\documentclass[lineno]{JFM-FLM_Au}

\lefttitle{I. Addison-Smith, I.A. Maia, B. Herrmann and A.V.G. Cavalieri}
\righttitle{Journal of Fluid Mechanics}

\title{Train yourself: self-compressing reduced-order models of turbulent flows}

\author{Ian Addison-Smith\aff{1}, Igor A. Maia\aff{2}, Benjamin Herrmann\aff{3,4} \and André V. G. Cavalieri\aff{2}}

\affiliation{\aff{1}Department of Mechanical Engineering, Universidad de Chile, Beauchef 851, Santiago, Chile
\aff{2}Divisão de Engenharia Aeronáutica, Instituto Tecnológico de Aeronáutica, São José dos Campos
12228-900, Brazil
\aff{3}Department of Mechanical and Metallurgical Engineering, Pontificia Universidad Católica de Chile, Av. Vicuña Mackenna 4860, Santiago, Chile
\aff{4}Department of Hydraulic and Environmental Engineering, Pontificia Universidad Católica de Chile, Av. Vicuña Mackenna 4860, Santiago, Chile
}

\corresau{Ian Addison-Smith, \email{ian.addisonsmith@ug.uchile.cl}}

\begin{document}
\maketitle

\begin{abstract}
Reduced-order models (ROMs) of turbulent flows based on Galerkin projection often require many degrees of freedom to resolve the dynamics of the turbulence, or simulation data to obtain an optimal modal basis. However, obtaining simulation data is computationally expensive, and the amount of data required to obtain a converged modal basis can increase this cost. Using the linearized Navier-Stokes equations, one can achieve spatial modes through the controllability and observability Gramians, which can yield a ROM without prior simulation data. In this work, we propose a self-compression of a ROM based on controllability modes, where the time series of the modal coefficients are leveraged to reduce the dimension of the ROM. In the self-compressed ROM (SCROM), we can maintain accurate first- and second-order statistics with respect to the DNS simulation, but in a further reduced dimension. The SCROM recovers spatial structures equivalent to proper orthogonal decomposition (POD) without relying on any simulation data, recombining spatial modes from linearized equations. This method leads to a novel ROM that can represent turbulence statistics in a data-free approach in a further reduced state space.

\end{abstract}



\section{Introduction}
\label{sec:headings}

In turbulent flows, obtaining reduced-order models (ROMs) from  Galerkin projections may require modal bases with a large number of modes to accurately represent the relevant dynamics \citep{rowley2017arfm}. Suitable modal bases for ROMs could be obtained from data using proper orthogonal decomposition (POD) (e.g. \citealt{noack2003jfm,smith2005jfm,aubry1988jfm,khoo2022jfm, sato2025jfm}), or from the linearized Navier-Stokes equations (e.g. \citealt{cavalieri2022prf, maia2025jfm, zong2025arxiv}). Statistical convergence of POD, as well as other data-driven modal decompositions~\citep{taira2017aiaa,herrmann2021jfm,baddoo2022prsa,schmid2022arfm,baddoo2023prsa}, requires large sequences of full flow field data snapshots. In particular, for wall-bounded turbulence, the streamwise and spanwise directions are homogeneous, so the modal bases are obtained from different streamwise and spanwise wavenumber pairs with a fixed truncation of eigenvectors to represent the wall-normal direction. Typically, the resulting models based on the linearized equations have a larger amount of degrees of freedom compared with the POD ones, but with the advantage that the resulting ROM does not require any prior data.

In this work, we present a framework to push for a further reduction in ROMs based on a modal bases from linearized equations. This self-compression leverages the time series of the modal coefficient of a ROM to solve an eigenvalue problem, where an energy optimal projection over a compressed space is obtained from the ROM. This approach, in comparison with common applications, utilizes data from the model itself, thereby maintaining the self-compression and ROM as data-free approaches. Therefore, one can significantly decrease the number of degrees of freedom of the ROM, while preserving a reasonable accuracy of first- and second-order statistics. Also, the computational cost can be significantly reduced in comparison with the ROM, which is relevant for control applications (e.g. \citealt{maia2025jfm})

In particular, we study a turbulent Couette flow with a Reynolds number ($\Rey = 500$), where we compare DNS simulations with a ROM built from controllability modes (same approach as \citealt{cavalieri2022prf}) and a self-compressed ROM (SCROM). Then we compare first- and second-order statistics of the three approaches to assess how much of the statistics are retained from the SCROM. Finally, we seek physical insight from the SCROM, obtaining spatial structures that compose the SCROM and the amount of wavenumber pairs that are selected in the model

\section{Methods}\label{sec:Methods}

\subsection{Reduced-order model formulation}

We consider plane Couette flow, wherein two infinite planes separated by a wall-normal distance $2h$ move in opposite directions with the same velocity magnitude $U_w$. All quantities are normalized by wall velocity and half-wall separation. The domain is defined in Cartesian coordinates by $(x,y,z)$ where $x$ is the streamwise, $y$ is the wall-normal and $z$ is the spanwise direction with dimensions $[0,2\pi) \times[-1,1] \times [0,\pi)$, boundary conditions $\mathbf{u}|_{y=\pm 1} = (\pm 1,0,0)$ in the upper and lower walls and periodic boundary conditions on streamwise and spanwise directions. The velocity field $\boldsymbol{u}$ is decomposed into a laminar solution $\boldsymbol{u}_0 = (y,0,0)$ plus fluctuations, $\boldsymbol{u}^{\prime}$, around this state, 
\begin{equation}
    \boldsymbol{u}(\boldsymbol{x},t) = \boldsymbol{u}_0(y) + \boldsymbol{u}^{\prime}(\boldsymbol{x},t),
\end{equation}
The governing equations for velocity fluctuations in an incompressible fluid are given by
\begin{align}
\frac{\partial \boldsymbol{u}' }{\partial t}
+ (\boldsymbol{u}_0 \cdot \boldsymbol{\nabla}) \boldsymbol{u}' 
+ (\boldsymbol{u}' \cdot \boldsymbol{\nabla}) \boldsymbol{u}_0& 
+ (\boldsymbol{u}' \cdot \boldsymbol{\nabla}) \boldsymbol{u}' 
= - \boldsymbol{\nabla} p 
+ \frac{1}{Re} \boldsymbol{\nabla}^2 \boldsymbol{u}', \label{eqn:navier-stokes} \\
\boldsymbol{\nabla} \cdot \boldsymbol{u}' &= 0.
\label{eqn:continuity}
\end{align}
where $p$ is the pressure and $Re=U_wh/\nu$ is the Reynolds number, with $\nu$ the kinematic viscosity of the fluid. Expanding the fluctuating velocity field in terms of a basis of divergence-free orthonormal modes, $\boldsymbol{\phi}_{j} (\boldsymbol{x})$, we can express the velocity as in~\citep{cavalieri2022prf}
\begin{equation}
    \boldsymbol{u}^{\prime} (\boldsymbol{x},t) = \sum_{j=1}^{N} a_j(t) \boldsymbol{\phi}_j(\boldsymbol{x}),
    \label{eqn:modal_velocity}
\end{equation}
where $a_j(t)$ are the corresponding modal expansion coefficients. Then, we substitute (\ref{eqn:modal_velocity}) into (\ref{eqn:navier-stokes}) and take an inner product with $\boldsymbol{\phi}_i$ to obtain the following equation for the dynamics of the coefficients
\begin{equation}
  \frac{\mathrm{d}a_i}{\mathrm{d}t} 
= \frac{1}{\Rey} \sum_{j=1}^N L_{ij} a_j
+ \sum_{j=1}^N \tilde{L}_{ij} a_j
+ \sum_{j=1}^N \sum_{k=1}^N Q_{ijk} a_j a_k ,
  \label{eqn:ROM_eq}
\end{equation}
where
\begin{subeqnarray}
    L_{ij} &=& \langle \nabla^2 \boldsymbol{\phi}_j , \boldsymbol{\phi}_i \rangle, \\
    \tilde{L}_{ij} &=& - \langle ( \boldsymbol{\phi}_j \cdot \nabla \boldsymbol{u}_0 + \boldsymbol{u}_0 \cdot \nabla \boldsymbol{\phi}_j ), \boldsymbol{\phi}_i \rangle, \\
    Q_{ijk} &=& - \langle ( \boldsymbol{\phi}_j \cdot \nabla \boldsymbol{\phi}_k ), \boldsymbol{\phi}_i \rangle,
    \label{eqn:ROM_op}
\end{subeqnarray}
and the symbols $\left < \cdot \right>$ denote an inner product, which we define in physical space as
\begin{equation}
\langle \boldsymbol{f}, \boldsymbol{g} \rangle = \frac{1}{4\pi^2} 
\int_{0}^{\pi} \int_{-1}^{1} \int_{0}^{2\pi} \boldsymbol{g}(\boldsymbol{x})^T 
\boldsymbol{f}(\boldsymbol{x})\mathrm{d}x \mathrm{d}y \mathrm{d}z,
\end{equation}
for real vector fields $\boldsymbol{f}$ and $\boldsymbol{g}$, with $( \ )^T$ denoting the transpose. The linear terms, $L_{ij}$ and $\tilde{L}_{ij}$, represent the viscous term and the linear interaction with the laminar solutions, respectively, and the nonlinear term, $Q_{ijk}$, represents the quadratic nonlinear interactions between the different modes. The term corresponding to the projection of $\phi_i$ upon pressure term vanishes due to the zero-pressure gradient condition and the divergence-free property of the modes. 
Here, for the choice of the modes $\boldsymbol{\phi}_j$, we follow the approach described in \cite{cavalieri2022prf}, who followed the method of \cite{jovanovic2005jfm}, wherein POD modes of the stochastic response of the linearized Navier-Stokes equations, also known as controllability modes, are considered. The controllability Gramian is computed for different streamwise/spanwise wavenumber combinations $(k_x,k_z)$, through a Lyapunov equation. The resulting basis consists of the eigenvectors of the Gramian that are ranked by their corresponding eigenvalues in terms of sustained energy. For each wavenumber pair considered, $n_{\text{modes}}$ eigenvectors are retained in the basis.

\subsection{Self-compressing reduced-order model}

Given the timeseries of the modal coefficients $a_j(t)$ obtained in a simulation of a ROM, we can separate these coefficients by wavenumber combination and create the following snapshot matrix for $(k_x/\alpha,k_z/\beta)=(m_x,m_z)$,
\begin{equation}
    \mathsfbi{X}_{(m_x,m_z)} = \begin{bmatrix}
\boldsymbol{a}_{(m_x,m_z)}(t_1) & \boldsymbol{a}_{(m_x,m_z)}(t_2) & \cdots  & \boldsymbol{a}_{(m_x,m_z)} (t_{N_s})
\end{bmatrix},
\label{eqn:wavenumber_snapshots}
\end{equation}
where $(\alpha, \beta) = (2\pi/L_x, 2\pi/L_z)$ are the fundamental streamwise and spanwise wavenumber, $(k_x/\alpha,k_z/\beta)$ are the wavenumber pair, and $N_s$ is the total number of snapshots obtained in the simulation of the ROM. This results in a matrix of dimensions $(n_{\text{modes}},N_s)$ containing the total snapshots for a number of modal coefficients corresponding to a wavenumber pair. If we consider that the ROM covers $\left| k_x \right| < M_x, \left| k_z \right| < M_z$ different wavenumber combinations, we can build a global snapshot matrix, $\mathsfbi{X}$, considering an arbitrary order for where the block matrices $\mathsfbi{X}_{(m_x,m_z)}$ are placed
\begin{equation}
    \mathsfbi{X} = \begin{bmatrix}
\mathsfbi{X}_{(0,1)} &  &  & & \\
 &  \mathsfbi{X}_{(0,2)}&  & & \\
 &  & \ddots  & & \\
 &  &  & \mathsfbi{X}_{(M_x,M_z)} &\\
 &  &  & &\mathsfbi{X}_{(0,0)}
\end{bmatrix}.
\label{eqn:full_snapshots}
\end{equation}

The inner product of the velocity field can be simplified with the orthogonality of the modes $\left \langle  \boldsymbol{\phi}_j, \boldsymbol{\phi}_k \right \rangle = \delta_{jk} $ as follows for arbitrary velocity decomposition $\boldsymbol{u}$ and $\boldsymbol{v}$
\begin{eqnarray}
    \langle \boldsymbol{u}, \boldsymbol{v} \rangle  &=& \sum_{k=1}^{N}\sum_{j=1}^{N} a_j(t) b_k(t)\left \langle  \boldsymbol{\phi}_j(\boldsymbol{x}), \boldsymbol{\phi}_k(\boldsymbol{x}) \right \rangle, \nonumber\\
    &=& \boldsymbol{a}^T\boldsymbol{b}. \label{eqn:norm_velocity}
\end{eqnarray}
If we consider $\boldsymbol{v}=\boldsymbol{u}$, then the inner product is $\langle \boldsymbol{u}, \boldsymbol{u} \rangle = \boldsymbol{a}^T\boldsymbol{a}$. This inner product is equivalent to the summation of the squared values of the modal coefficients. Also, a covariance matrix of $\mathsfbi{X}$ is defined as
\begin{equation}
        \mathsfbi{R}  = \mathsfbi{X}\mathsfbi{X}^T.
    \label{eqn:covariance_matrix}
\end{equation}
Then, given the snapshot matrix $\mathsfbi{X}$, we can search for the linear combinations of modes for every wavenumber combination $(k_x/\alpha,k_z/\beta)$ that can optimally represent the energy in the ROM simulation. This can be achieved by formulating an eigenvalue problem for the matrix $\mathsfbi{R}$
\begin{equation}
    \mathsfbi{R}  \mathsfbi{T}  = \mathsfbi{T}  \mathsfbi{\Lambda} .
    \label{eqn:eigenvalue_problem}
\end{equation}
The diagonal matrix of eigenvalues, $\mathsfbi{\Lambda}$, orders hierarchically by energy the combination of modes in each wavenumber combination,  represented by each column of the matrix $\mathsfbi{T}$. This method resembles proper orthogonal decomposition (POD) \citep{Holmes_turbulence_book} as velocity fluctuations are represented through the modal coefficients; however, this method does not rely on any previous simulation or experimental data. A further compression of the ROM system can be obtained by truncating the total energy to a threshold $r$. This threshold then sets a compressed number of modes, $n$, required to achieve the energy reduction. We can then define a truncated eigendecomposition
%
\begin{equation}
\mathsfbi{R} \mathsfbi{T} \approx \mathsfbi{T}^{(n)} \Lambda^{(n)},
\label{eqn:eigenvalue_truncation}
\end{equation}
of corresponding rank $n$. Using the truncated matrix $\mathsfbi{T}^{(n)}$ and the pseudo-inverse ${\mathsfbi{T}^{(n)}}^{\dagger}$, we can transform the reduced modal coefficients to a compressed form of modal coefficients with the following equations
\begin{subequations}
\begin{equation}
a_i(t) =  \sum_{p=1}^n T_{ip}^{(n)} b_p(t),
\label{eqn:rrom_to_rom}
\end{equation}
\begin{equation}
b_p^{(n)}(t) = \sum_{i=1}^N {T^{(n)}}^{\dagger}_{pi} a_i(t).
\label{eqn:rom_to_rrom}
\end{equation}
\end{subequations}
A self-compressing ROM (SCROM) can be obtained using \ref{eqn:rom_to_rrom} in the time evolution of the modal coefficients, resulting in an equation for the time evolution of the compressed modal coefficients $\boldsymbol{b}^{(n)}(t)$
\begin{equation}
  \frac{\mathrm{d}b_p^{(n)}}{\mathrm{d}t} 
= \frac{1}{\Rey} \sum_{q=1}^n L_{pq}^{(n)} b_q
+ \sum_{q=1}^n \tilde{L}_{pq}^{(n)} b_q
+ \sum_{q=1}^n \sum_{r=1}^n Q_{pqr}^{(n)} b_q b_r,
  \label{eqn:RROM_eq}
\end{equation}
where
\begin{subeqnarray}
    L_{pq}^{(n)} &=& \sum_{i=1}^N \sum_{j=1}^N {T^{(n)}}^{\dagger}_{pi} L_{ij} T_{jq}^{(n)} \\
    \tilde{L}_{pq}^{(n)} &=& \sum_{i=1}^N \sum_{j=1}^N {T^{(n)}}^{\dagger}_{pi} \tilde{L}_{ij} T_{jq}^{(n)} \\
    Q_{pqr}^{(n)} &=& \sum_{i=1}^N \sum_{j=1}^N \sum_{k=1}^N {T^{(n)}}^{\dagger}_{pi} Q_{ijk} T_{jq}^{(n)} T_{kr}^{(n)}.
    \label{eqn:RROM_op}
\end{subeqnarray}

The system of equations \ref{eqn:RROM_eq} describes the time evolution of the compressed coefficients of the SCROM, where the compression of the vector state to $\boldsymbol{b}^{(n)}(t)$ with $n < N$ can achieve significant computational savings while maintaining, as will be shown shortly, a reasonable accuracy in turbulent statistics.

\subsection{Numerical implementation}

As mentioned above, the modal basis, $\boldsymbol{\phi}_i$, consisted of eigenfunctions of the linearized Navier-Stokes system forced with white noise in space and time, as in \cite{farrell1993pof,jovanovic2005jfm,herrmann2023jfm}. Following the framework of \cite{cavalieri2022prf}, we obtain orthonormal velocity modes using as a base flow the laminar solution $\boldsymbol{U}(y) = (y,0,0)$ and $\Rey=100$ for different wavenumber combinations $(k_x/\alpha,k_z/\beta)$. Controllability modes for wavenumber combinations $\left| k_x/\alpha \right| < 2, \left| k_z/\beta \right| < 2$ are computed, and $n_{\text{modes}}=24$ modes are retained for each wavenumber pair. In order to deal with real modes only, we decompose each eigenfunction into its real and imaginary parts, which leads to a pair of modes separated by a phase shift of $\pi/2$. This results in a total amount of $600$ modes. Modes with $(-k_x,-k_z)$ are dropped, as they are simply complex conjugates of $(k_x,k_z)$, so only positive $k_x$ is considered. In the case of $k_x = k_z =0$ corresponding to mean-flow modes, we use Stokes modes (eigenfunction of a linear viscous diffusion problem) as in other works (e.g. \citealt{waleffe1997pof,cavalieri2022prf,maia2025jfm})

We define a mesh $(N_x,N_y,N_z)=(10,65,10)$ using equidistant points in streamwise and spanwise directions. The wall-normal direction is discretised with Chebyshev polynomials, whereas the homogeneous directions are discretised with Fourier modes. Using an initial condition of $a_i \in (0,0.1)$, the systems in \ref{eqn:ROM_eq} and  \ref{eqn:RROM_eq} are integrated up to 3000 non-dimensional time units ($tU_w/h$) with a 0.5 timestep and 0.125 timestep respectively, using the \texttt{ode45} function in MATLAB. Statistics are collected from the time series excluding the first 100 time units. A reference DNS simulation is obtained using the Channelflow spectral code \citep{gibson2008jfm,gibson2014chflow}. For the DNS, the grid posesses $(N_x,N_y,N_z)=(32,65,32)$ points, with a factor of $3/2$ factor in $N_x$ and $N_z$ for de-aliasing, and statistics are also collected after a transient of 100 time units. 

The ROM described in equation \ref{eqn:ROM_eq}, is first run in order to store the snapshot matrices and compute the truncated matrix, $\mathsfbi{T}^{(n)}$, which is necessary to run the SCROM, \ref{eqn:RROM_eq}. The snapshot matrix is also built excluding the first 100 time units. Snapshots of the same wavenumber can be related to the real part or the imaginary part of the eigenmode, so we data augment the snapshots matrix with data from the same wavenumber but real/imaginary part, respectively. Using this idea, the amount of snapshots matrix in equation \ref{eqn:wavenumber_snapshots} is duplicated. 

To quantify the accuracy of different reduction levels, an error metric is defined based on the $L^{2}$ norm of the difference in first-order and second-order statistics with respect to the ROM, as defined in \cite{maia2024tcfd}. For instance, the error in the representation of the $\overline{u'u'}$ component of the Reynolds tensor is given by
\begin{equation}
    \varepsilon_{(u'u')} = \frac{\left\| \overline{(u'u')}_{\text{(ROM)}} - \overline{(u'u')}_{\text{(SCROM)}} \right\|_2}{\left\| \overline{(u'u')}_{\text{(ROM)}} \right\|_2},
    \label{eqn:err_stats}
\end{equation}
where $\overline{(\cdot)}$ denotes averaging in the streamwise and spanwise directions and in time and the subscript indicates the model to compare (ROM or SCROM). The $L^2$ norm is equivalent to the inner product as $\left\| \boldsymbol{u} \right\|_2 = \sqrt{\langle \boldsymbol{u}, \boldsymbol{u} \rangle}  $ 

\section{Results}\label{sec:results}

Following the compression procedure shown in section \ref{sec:Methods}, we explore different levels of compression,  corresponding to different thresholds, $r$, computed from the eigenvalue problem \ref{eqn:eigenvalue_problem}.

\begin{figure}
  \centerline{\includegraphics[width=1\textwidth]{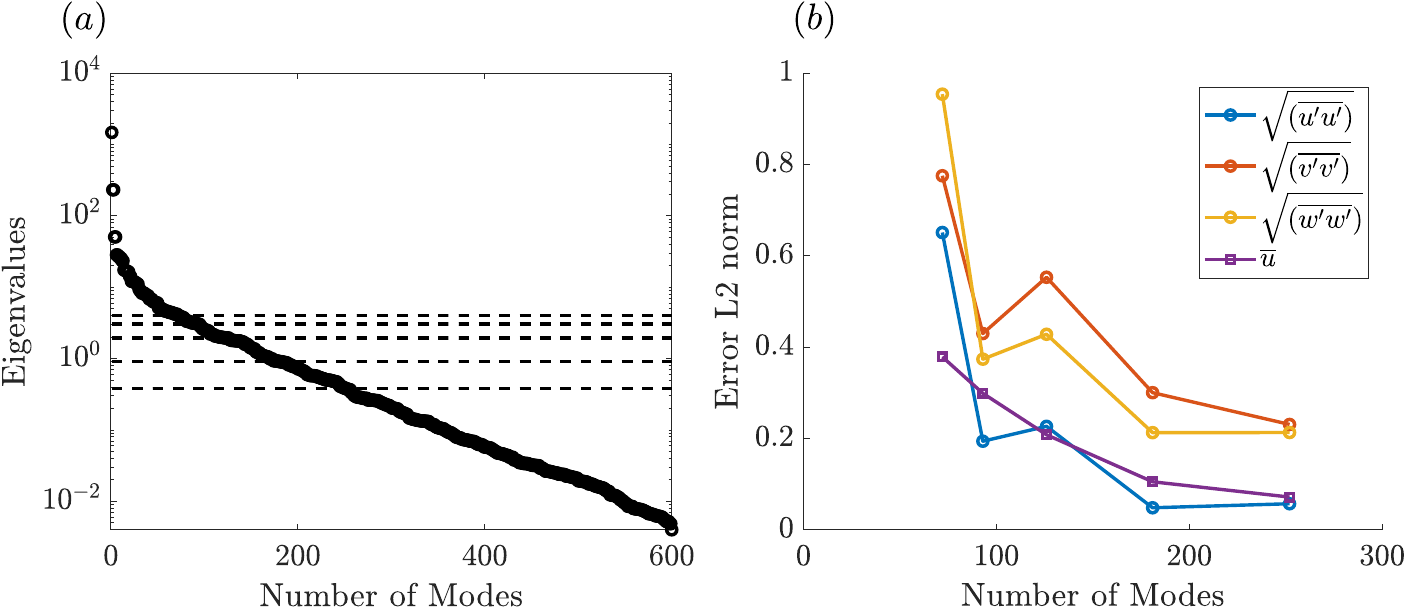}}
  \caption{(a) Eigenvalues of the matrix $\mathsfbi{R}$, different thresholds are shown by horizontal lines in descending order $r=0.9,0.925,0.95,0.975,0.99$ which correspond to $n=72,93,126,181,252$ retained modes, respectively. (b) Error of the first and second-order statistics for the same thresholds as (a).}
\label{fig:fig1}
\end{figure}

\subsection{Statistics of the self-compressed form}

The eigenvalues corresponding to the covariance matrix $\mathsfbi{R}$ and the different thresholds are shown in figure \ref{fig:fig1}(a), where the various thresholds are indicated by horizontal lines that intersect the eigenvalues. Figure \ref{fig:fig1}(b) shows the corresponding errors in the flow statistics computed from SCROMs with different number of modes with respect to the ROM. The overall behavior shows a monotonic decay of the error in first-order statistics; however, the second-order statistics exhibit non-monotonic behavior with increasing number of modes. Non-monotonic convergence has also been reported in DNS with coarse grids \citep{meyers2007pof,rasam2011jot}.

\begin{figure}
  \centerline{\includegraphics[width=1\textwidth]{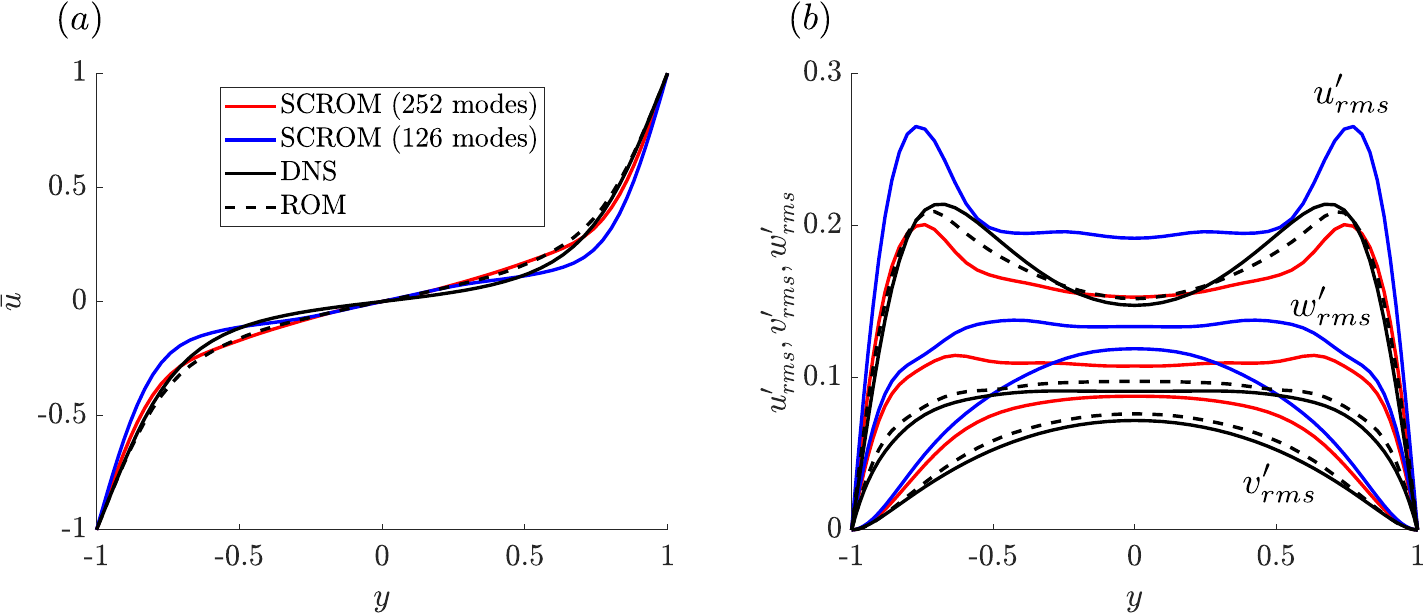}}
  \captionsetup{style=capcenter}
  \caption{(a) First-order statistics and (b) second-order statistics for different thresholds of the SCROM}
\label{fig:fig2}
\end{figure}

A further comparison of the statistics is displayed in figure \ref{fig:fig2} for two of those cases. In \ref{fig:fig2}(a,b), first and second-order statistics of ROM, SCROM and DNS are compared. A threshold of $r=0.99$ (or $n=252$) in the SCROM produces a reasonable agreement with respect to the ROM, while retaining less than half the number of modes used in the original system. A more severe truncation of $r=0.95$ (or $n=126$) considerably increases the error in second-order statistics, especially for the $v$ and $w$ components. However, even the smallest SCROM produces a mean flow in reasonable agreement with the ROM and the DNS, which can be seen in figure \ref{fig:fig2}(a).

\subsection{Interpretation of the self-compressed form}

\begin{figure}
  \centerline{\includegraphics[width=1\textwidth]{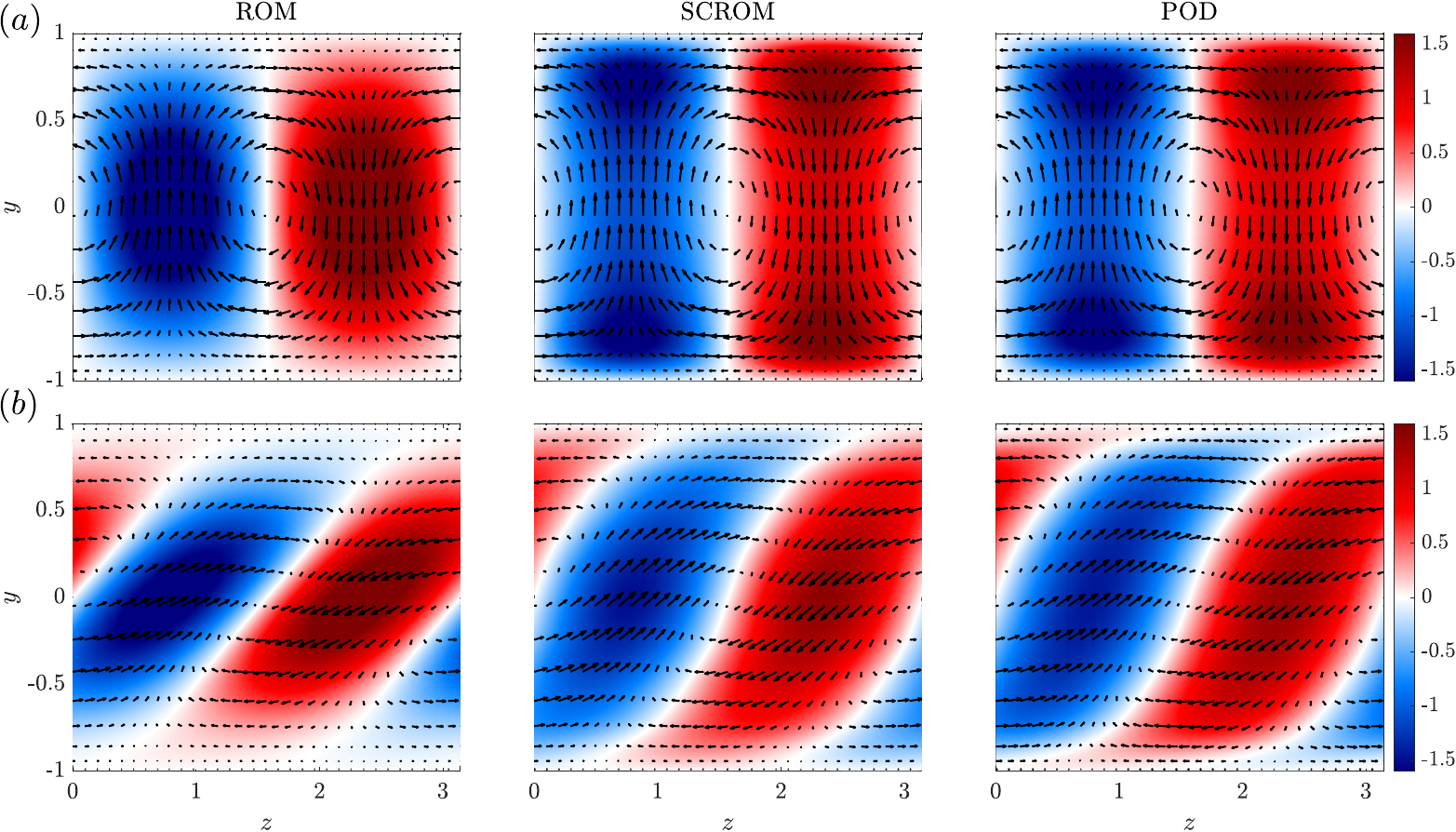}}
  \captionsetup{style=capcenter}
  \caption{Visualization of the leading modes in the ROM, SCROM and POD bases for wavenumber pairs $(k_x/\alpha,k_z/\beta)=(0,1)$ in (a) and $(k_x/\alpha,k_z/\beta)= (1,1)$ in (b). The positive and negative contours indicate streamwise component, whereas the arrows indicates the wall-normal and spanwise component}
\label{fig:fig3}
\end{figure}

The basis of the SCROMs stems from an optimal compression (in terms of an $L_2$-norm) of the original controllability modes, obtained from linearized equations for each wavenumber pair, as shown in equations \ref{eqn:full_snapshots} and \ref{eqn:eigenvalue_problem}. The compression produces modified modes, which are illustrated in figure \ref{fig:fig3} for two pairs $(k_x/\alpha,k_z/\beta)=(0,1)$ and $(k_x/\alpha,k_z/\beta)=(1,1)$. The modified modes of the SCROM describe flow structures that are quite similar to POD modes computed from DNS data, but which are obtained without any prior knowledge of the latter. This modal basis leverages the nonlinear structure of the ROM, represented by $Q_{ijk}$, to recover structures that consider the influence of nonlinearity, that is not present in the controllability modes of the ROM.

\begin{figure}
  \centerline{\includegraphics[width=1\textwidth]{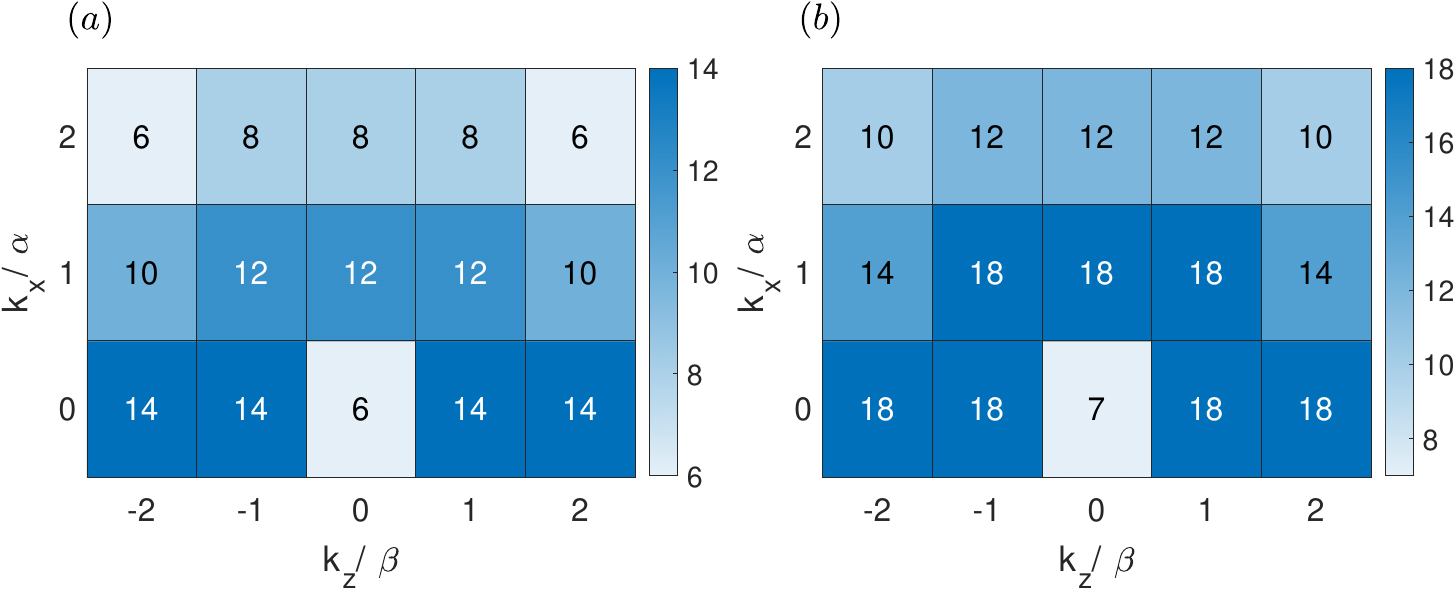}}
  \captionsetup{style=capcenter}
  \caption{Number of modes retained in the SCROM basis for each wavenumbers pair $(k_x/\alpha,k_z/\beta)$: (a) threshold of $r = 0.95$; (b) threshold of $r=0.975$.}
\label{fig:fig4}
\end{figure}

The compression procedure laid out here also quantifies the energetic relevance of structures at each wavenumber pair, as the number of modes retained for each $(k_x, k_z)$ is, in principle, different. This is illustrated in figure \ref{fig:fig4}(a,b). The number of retained modes for each wavenumber pair is dependent on the energy threshold $r$; also, increasing $r$ (or the energy captured by the reduction) does not imply a proportional increase in the number of modes for each wavenumber pair, as can be seen by comparing figures \ref{fig:fig4}(a,b).

\section{Conclusions}\label{sec:conclusions}

A key aspect of reduced-order modeling is the ability to describe the dynamics of turbulent flow in a lower-dimensional state space. Relying on an equation-based approach, where modes arising from linearized equations are shown to give a model with reasonable statistics \citep{cavalieri2022prf}. Here, we present a data-free approach to enhance the model reduction, leveraging the equations and symmetries in a parallel flow setup. Using different thresholds of energy as truncations, we derive different self-compressed reduced models that have good agreement with the ROM and the DNS simulation. Specifically a model with $n=252$ ($r=0.99$) can retain the first and second-order statistics of the turbulent flow, which corresponds to approximately a reduction of a half of modes in comparison with the ROM.

Further physical interpretation of the compressions sheds light on the spatial structure of the modes and the selection of different wavenumber pairs. In the work of \cite{cavalieri2022prf}, they demonstrated that the accuracy of the ROM depends on the careful selection of different wavenumber pairs and the number of modes that correspond to each pair. Additionally, the modal bases play a crucial role \citep{zong2025arxiv}, where the accuracy of the turbulence statistics for a fixed wavenumber pair model depends on the linearized equations from which the modal basis is obtained. Here, after the compression step, we arrive at structures that are similar to POD modes without requiring prior data. These structures optimally represent the turbulence statistics in a parallel flow configuration \citep{zong2025arxiv}, in contrast to other modal bases. Also, the compression shows that the choices of modes by wavenumber pair is not fixed, where different structures weigh differently in the SCROM. 

This work opens new horizons for reduced-order modeling applications, specifically where the current dimension is still large enough, but the accuracy of statistics is sufficient. Specifically, these models can be applied, for example, to turbulence control \citep{maia2025jfm}, where turbulence can be suppressed using a specific forcing, the computation of invariant solutions of the turbulent state \citep{mccormack2024jfm} and push ROMs for turbulent flows with higher Reynolds number. This self-compression methodology enables reducing the computational cost for different applications while maintaining a good agreement in the turbulence state. Additionally, it provides a connection between purely data-driven methods, such as POD-Galerkin, and equation-based methods as presented in this work.

\begin{bmhead}[Acknowledgements]
We gratefully acknowledge Y. Hwang for his helpful comments and insightful discussions.
\end{bmhead}

\begin{bmhead}[Funding]
Ian Addison-Smith acknowledges the financial support received from CAPES (Coordenação de Aperfeiçoamento de Pessoal de Nível Superior -
Brasil (CAPES) - Finance Code 001), under Move La America program, and ANID under scholarship ANID-Subdirección de Capital Humano/Magíster Nacional/2024-22241812. This work was also funded by ANID Fondecyt project 1250693.
\end{bmhead}

\begin{bmhead}[Declaration of interests]
The authors report no conflict of interest.
\end{bmhead}

\bibliographystyle{jfm}

\end{document}